\documentclass[12pt,oneside]{article}
\input psfig.sty
\begin{document}
\thispagestyle{empty}
\setcounter{page}{0}
\renewcommand{\theequation}{\thesection.\arabic{equation}}

\def\laq{\raise 0.4ex\hbox{$<$}\kern -0.8em\lower 0.62
ex\hbox{$\sim$}}
\def\gaq{\raise 0.4ex\hbox{$>$}\kern -0.7em\lower 0.62ex\hbox{$\sim$}}

\def\baselinestretch{1.4} \def\tdot#1{{\buildrel{\ldots}\over{#1}}}
\setlength{\oddsidemargin}{0.0cm} \setlength{\textwidth}{16.5cm}
\setlength{\topmargin}{-.9cm} \setlength{\textheight}{22.5cm}%

\font\tenbb=msbm10
\font\sevenbb=msbm7
\font\fivebb=msbm5
\newfam\bbfam
\textfont\bbfam=\tenbb \scriptfont\bbfam=\sevenbb
\scriptscriptfont\bbfam=\fivebb

\def\bb{\fam\bbfam}
\def\Rb{{\bb R}}

\newcommand{\vrsmall}{\vrule width 0pt height 18pt depth 15pt}
\newcommand{\vrbig}{\vrule width 0pt height 27pt depth 15pt}
\newcommand{\vrh}{\vrule width 0pt height -40pt depth 0pt}
\newcommand{\pa}{\partial} 
\newcommand{\co}{\nabla}
\newcommand{\vphi}{\varphi} 
\newcommand{\sigmap}{\sigma^\prime}
\newcommand{\taup}{\tau^\prime}
\newcommand{\beq}{\begin{equation}} 
\newcommand{\eeq}{\end{equation}}
\newcommand{\bea}{\begin{eqnarray}} 
\newcommand{\eea}{\end{eqnarray}}
\newcommand{\beam}{\begin{mathletters}} 
\newcommand{\eeam}{\end{mathletters}}
\newcommand{\var}{\rm Var} 
\newcommand{\ophi}{\overline{\phi}} 
\newcommand{\obeta}{\overline{\beta}} 
\newcommand{\om}{\overline{m}} 
\newcommand{\opsi}{\overline{\psi}} 
\def\p{\partial}
\def\mp{m_{\rm pl}}
\def\lap{\lower.5ex\hbox{$\; \buildrel < \over \sim \;$}}
\def\gap{\lower.5ex\hbox{$\; \buildrel > \over \sim \;$}}
\def\lb{\langle}
\def\rb{\rangle}
\def\fv{{\bf f}}
\def\vp{\varphi}
\font\tenbb=msbm10
\font\sevenbb=msbm7
\font\fivebb=msbm5
\newfam\bbfam
\textfont\bbfam=\tenbb \scriptfont\bbfam=\sevenbb
\scriptscriptfont\bbfam=\fivebb
\def\bb{\fam\bbfam}
\def\Rb{{\bb R}}

{\hfill{IHES/P/02/22}} 


\vspace{2cm}

\begin{center}
{\bf STRING COSMOLOGY AND CHAOS}

\vspace{1.4cm}

Thibault DAMOUR

\vspace{.2cm}

{\em Institut des Hautes Etudes Scientifiques, 35 route de Chartres,}\\

{\em 91440 Bures-sur-Yvette, France} \\
\end{center}

\vspace{-.1cm}

\centerline{{\tt damour@ihes.fr}}

\vspace{1cm}

\centerline{ABSTRACT}

\vspace{- 4 mm}  

\begin{quote}\small
We briefly review three aspects of string cosmology: (1) the ``stochastic'' approach to the
pre-big bang scenario, (2) the presence of chaos in the generic cosmological solutions of the
tree-level low-energy effective actions coming out of string theory, and (3) the remarkable
link between the latter chaos and the Weyl groups of some hyperbolic Kac-Moody algebras.
Talk given at the Francqui Colloquium ``Strings and Gravity: Tying the Forces Together''
(Brussels, October 2001).
\end{quote}

\baselineskip18pt

\newpage

\setcounter{equation}{0}
\section{Introduction}
A striking prediction of string theory is that its ``gravitational sector'' is 
richer than that of General Relativity: it contains several {\it a priori} massless
fields in addition to the Einstein graviton (see \cite{GSW, polchinski} for  reviews). 
We consider that these fields (notably the dilaton $\phi$, as well as possibly some 
of the other stringy partners of the graviton) represent an interesting 
prediction  whose possible existence 
should be taken seriously, and whose observable consequences should be 
carefully studied. 
Of course, tests of General Relativity  put severe constraints on such fields, and
notably on the dilaton. The simplest way to recover General Relativity at late times
is to assume \cite{TV} that $\phi$ gets a mass from supersymmetry-breaking non-perturbative 
effects. Another possibility might be 
to use the string-loop modifications of the dilaton couplings for driving 
$\phi$ toward a special value where it decouples from matter \cite{DP}. 
[Recently Ref.\cite{DPV} has explored the phenomenological consequences of
a version of this cosmological decoupling scenario where the special value of the
dilaton corresponds to infinite bare string coupling.]
These alternatives do not rule out the possibility
that the dilaton may have had an important r\^ole  in the previous  history of the
universe. Early cosmology stands out as a particularly 
interesting arena where to study the dynamical effects of the dilaton and of the
other stringy partners of the graviton.
In this contribution, we wish to discuss two separate attempts at exploring the 
cosmological consequences of the richer stringy gravitational sector.

In a first part, we shall briefly review one facet of the pre-big bang (PBB) model 
\cite{1,PBB,copeland} of early string cosmology: the  ``stochastic'' approach to the
problem of initial conditions \cite{BDV}.
 Then we shall summarize some recent work,
done in collaboration with Marc Henneaux \cite{DH1,DH2,DH3,DHJN,DHN},
 which discovered the generic presence
of a chaotic behaviour in string cosmology, and the link of this chaos with the
Weyl groups of remarkable hyperbolic Kac-Moody algebras ($E_{10}$, $BE_{10}$, ...).

\setcounter{equation}{0}

\section{Stochastic pre-big bang}

A series of papers \cite{1,PBB,BDV,copeland} has developed the so-called pre-big bang (PBB) model,
in which the dilaton plays a key dynamical r\^ole.
One of the key ideas of this scenario is to use 
the kinetic energy of the dilaton to drive a period of inflation of the 
universe. The motivation is that the presence of a (tree-level 
coupled) dilaton essentially destroys \cite{BS} the usual inflationary mechanism: 
 instead of driving an exponential inflationary expansion, a 
(nearly) constant vacuum energy drives the string coupling $g = e^{\phi / 2}$ 
towards small values, thereby causing the universe to expand only as a small 
power of time. If one takes seriously the existence of the dilaton, the PBB 
idea of a dilaton-driven inflation offers itself as one of the very few natural 
ways of reconciling string theory and inflation.

Let us first recall that, within the PBB scenario, the inflation driven by the kinetic 
energy of $\phi$ forces both the coupling and 
the curvature to {\it grow} during inflation \cite{1}. This suggests that the 
initial state must be very perturbative in two respects: i) it must have
very small initial curvatures (derivatives) in string units and, ii) it must exhibit 
a tiny initial coupling $g_i = e^{\phi_i /2} \ll 1$.  
In conclusion, dilaton-driven inflation must start
from a regime in which the tree-level low-energy approximation 
to string theory is extremely accurate, something we may call an asymptotically trivial
``vacuum'' state.

The first papers on the PBB scenario were assuming that this initial ``trivial'' 
vacuum state was, in addition, very symmetric (spatially homogeneous). 
Several authors \cite{TW}, \cite{KLB}
criticized this ``fine tuning'' of the initial conditions. This led \cite{BDV}
to develop a ``stochastic'' version of the PBB scenario that we wish to explain.
We shall first follow \cite{BDV} in assuming that the set of considered string 
vacua are already compactified to
 four dimensions and are truncated to the 
gravi-dilaton sector (antisymmetric tensors and moduli being set to zero).
As we shall see in the next section, this assumption, mostly chosen ``for simplicity's sake'',
turns out to modify in a drastic way the qualitative behaviour of the general (tree-level)
stringy cosmological solution near the big-crunch/big-bang singularity. This lesson should
be kept in mind when exploring other simplified models of a possible 
big-crunch/big-bang transition \cite{KOSST,S02,N02}.

Within this simplified framework, the set of all perturbative string vacua coincides 
with the generic solutions of the tree-level low-energy effective action 

\beq
S_s = \frac{1}{\alpha^\prime} \int d^4 x \, \sqrt{G} \, e^{-\phi} [R(G) + G^{\mu 
\nu} \partial_{\mu} \phi \, \partial_{\nu} \phi] \,, 
\label{eqn1.1}
\eeq
where we have denoted by $G_{\mu \nu}$ the string-frame ($\sigma$-model) metric.
The generic solution is parametrized by 6 functions of three variables.  These 
functions can be thought of classically as describing the two helicity$-2$ 
modes of gravitational waves, plus the helicity$-0$ mode of dilatonic waves. 
The idea is then to envisage, as initial state, 
the most general past-trivial classical solution of (\ref{eqn1.1}), i.e. an 
arbitrary ensemble of incoming gravitational and dilatonic waves. 
 
The main point of \cite{BDV} was to show how such a stochastic bath of classical 
incoming waves (devoid of any ordinary matter) can evolve into our rich,
complex, expanding universe. The basic mechanism  for turning such a 
trivial, inhomogeneous and anisotropic, initial state into a Friedmann-like 
cosmological universe is {\it gravitational instability}
 (and quantum particle production as far as
heating up the universe is concerned \cite{PBB}). When the 
initial waves satisfy a certain (dimensionless) strength criterion, they 
collapse (when viewed in the Einstein conformal frame) under their own weight. 
As discussed below, when viewed in the (physically most appropriate) string conformal frame,
a fraction of these collapses (the ones where  $\sum_a \alpha_a < 0$) 
leads to the local birth of  baby inflationary universes 
blistering off the initial vacuum. Assuming that the dilaton-driven (power-law)
inflation is somehow converted into a hot big bang,
 one  then expects each of these ballooning patches 
of space to evolve into a quasi-closed Friedmann universe
\footnote{This picture of baby universes created by gravitational 
collapse is reminiscent of earlier proposals \cite{FMM}, \cite{MB}, 
\cite{Smolin}.}.

Though the physical interpretation of such a model is best made in terms of the 
original string (or $\sigma$-model) metric $G_{\mu \nu}$ appearing in 
Eq.~(\ref{eqn1.1}), it is technically convenient to work with the 
conformally related Einstein metric
\beq
g_{\mu \nu} = e^{-(\phi -\phi_{\rm now})} \, G_{\mu \nu} \, , 
\quad \quad 16 \pi G = \alpha^\prime \, e^{\phi_{\rm now}}\,. 
\label{eqn2.1}
\eeq
In terms of the Einstein metric $g_{\mu \nu}$, the low-energy tree-level string 
effective action (\ref{eqn1.1}) reads (we  set, here,  $16 \pi G = 1$)
\beq
\label{1.1}
S = \int d^4 x \,\sqrt{g}\,
\left [ {R} - \frac{1}{2}\pa_\mu \phi\, \pa^\mu \phi \right ]\,.
\eeq
The corresponding classical field equations are
\beq
R_{\mu \nu} = \frac{1}{2} \,\partial_{\mu} \phi \, \partial_{\nu} \phi \, ,  
\label{eq1.4} 
\eeq
\beq
\nabla^{\mu} \, \nabla_{\mu} \, \phi = 0 \, .
\label{eqn2.3}
\eeq
As explained in \cite{BDV} a generic solution of these 
classical field equations admitting an asymptotically trivial past, i.e. a
generic stringy ``in state'', can be
described as a superposition of incoming wave packets of gravitational 
and dilatonic fields. This ``in state" can be nicely parametrized by three 
asymptotic ingoing, dimensionless ``news'' functions $N(v,\theta,\varphi)$, $N_+ 
(v,\theta,\varphi)$, $N_{\times} (v,\theta,\varphi)$.  
 When  all the news stay always significantly below 1, this ``in state'' will 
evolve into a similar trivial ``out state'' made of outgoing wave packets. On the 
other hand, when the news functions reach values of order 1, and more precisely 
when some global measure (discussed in detail in \cite{BDV}) 
of the variation  of the news functions
 exceeds some critical value of order unity, the ``in state'' will 
become gravitationally unstable during its evolution and  will give 
birth to one or several black holes, i.e. one or several singularities hidden 
behind outgoing null surfaces (event horizons). Seen from the outside of these 
black holes, the ``out" string vacuum will finally look, like the ``in" one, as a 
superposition of outgoing waves.
However, the story is very different if we 
look inside these black holes and shift back to the physically more appropriate 
string conformal frame.

It is at this point that the ``simplification'' of considering only the 
Einstein-dilaton system (\ref{1.1}) plays a particularly crucial role.
Indeed, the  work of Belinsky, Khalatnikov and Lifshitz \cite{BKL} has 
shown that the qualitative behaviour of the fields near a generic 
cosmological singularity depended very much on the ``menu'' of fields
present in the theory. In particular, Belinsky and Khalatnikov \cite{BK}
found that (in any dimension) the Einstein-dilaton system admitted a simple
``Kasner-like'' monotonic behaviour near a space-like singularity.
[See \cite{AR00}, \cite{DHRW} for  mathematical proofs of this result.]
This contrasts with, for instance, the generic behaviour of the pure
Einstein system (in dimension $ D < 11$), which exhibits a very 
complicated behaviour comprising an infinite number of shorter and shorter
``oscillations'' near a singularity (see below). 

In the Einstein frame, one then
finds that the the Einstein-dilaton system leads to a monotonic, power-law-type,
``collapse'' near the big-crunch singularity. This 
monotonic behaviour is technically described (in any dimension, and in suitable,
Gaussian coordinates) by a spatially inhomogeneous version of the Kasner solution. 
We shall
give in Eq.(\ref{eq2}) below the Einstein-frame expression of this Kasner-like solution. 
Let us indicate here its string-frame version:
\beq
\label{strkasn1}
ds_S^2 \sim -dt_S^2 + \sum_{a=1}^{D-1} (-t_S)^{2\alpha_a (x)} 
(E_i^a (x) \, dx^i)^2 \, ,
\eeq
\beq
\label{strkasn2}
\phi (x,t) = \phi (x,0) + \sigma (x) \log (-t_S) \, ,
\eeq
where the spatial string-frame $(D-1)$-bein $E_i^a (x)$ is proportional to the  
Einstein-frame $(D-1)$-bein $e_i^a (x)$ of Eq.(\ref{eq2}).
The constraints that must be satisfied by  the ``string-frame Kasner exponents'' 
read ($1 \leq a \leq D-1$)
\beq
\label{strrel}
\sum_{a=1}^{D-1} \alpha_a^2 = 1 \,, \quad \quad 
\sigma = \left( \sum_{a=1}^{D-1} \alpha_a \right) - 1 \,.
\eeq

The link between the (spatially varying) string-frame exponents $\alpha_a(x)$,
and the Einstein-frame ones $p_a(x), p_{\varphi}(x)$ (introduced below) reads
\beq
\label{strein}
p_a = \frac{\alpha_a(D-2) - {\sigma}}{D-2-{\sigma}}\,,
\quad \quad 
p_{\varphi} = \frac{\sqrt{ (D-2)}\,\sigma}{D-2 - {\sigma}}\,,
\eeq
where $\sigma \equiv (\sum_a \alpha_a) - 1$.

When described in the string frame, the 
Einstein-frame collapse towards a space-like singularity will represent, if 
$\phi$ grows fast enough toward the singularity (more precisely if 
$\sum_a \alpha_a < 0$, so that the volume in string units grows) 
a ``super-inflationary'' expansion of space (i.e. such that the volume grows like
a negative power of $-t_S$, as $t_S \to 0^-$). 
The picture is therefore that inside each black hole, the regions near the 
singularity where $\phi$ grows sufficiently fast will blister off the initial 
trivial vacuum as many separate pre-Big Bangs. The PBB scenario assumes that these
inflating pre-big bang patches (which head toward a singularity at $t_S = 0$, where
$\phi$ and the curvature blow up) make a ``graceful'' transition toward a (decelerated)
Friedmann-Lema\^{\i}tre hot big bang state.
These inflating patches are 
surrounded by non-inflating, or deflating  
patches, and therefore globally look approximately like
closed Friedmann-Lema\^{\i}tre hot universes. 
[See Figures 1 and 2 of \cite{BDV} for sketches of this picture.]
One expects such quasi-closed universes to recollapse 
in a finite, though very long, time (which is 
consistent with the fact that, seen from the outside, 
the black holes therein contained must evaporate in a finite time).

This picture has been firmed up by the detailed analysis of the spherically symmetric
Einstein-scalar system in \cite{BDV}. However, the weakest part of the entire 
PBB scenario is the conjectural assumption that the above power-law big-crunch 
behaviour, with a locally growing string coupling, can be ``halted'' by 
various non-perturbative effects (particle creation, string loops, ...) and
``reversed'' into a decelerated Friedman-like hot big-bang. For references on this
``bounce'' problem within the PBB scenario see \cite{PBB, copeland}, and for recent work
within some string-theory toy models see \cite{KOSST,S02,N02}. We shall, however, see
in the following section that taking into account all the (massless) fields entering
the low-energy action (and notably the Ramond-Ramond fields) drastically alters
the simple behaviour of the fields near a big-crunch singularity.

 \setcounter{equation}{0}
\section{Chaos in Superstring Cosmology}

 A crucial problem in
string theory is the problem of vacuum selection. It is reasonable to believe that this problem
can be solved only in the context of cosmology, by studying the time evolution of {\it generic},
inhomogeneous (non-SUSY) string vacua. We have seen in the previous section that the
generic (inhomogeneous) solution of the simple Einstein-dilaton 
system (\ref{eqn1.1}) displayed (especially when viewed in the string frame) 
a rather rich structure. Let us recall that Belinskii, Khalatnikov and Lifshitz (BKL) have 
discovered \cite{BKL} that the generic solution of the {\it four-dimensional} 
Einstein's vacuum equations had a much richer, and much more complex structure, characterized
by a non-monotonic, never ending oscillatory behaviour near a cosmological singularity.
The oscillatory approach toward the 
singularity has the character of a random process, whose 
chaotic nature has been intensively studied \cite{LLK}.
 [See \cite{berger} for a summary of the
evidence supporting the BKL conjectural picture.]
The qualitative behaviour of the generic solution near a cosmological singularity 
depends very much: (i) on the
field content of the system being considered, and (ii) on the spacetime dimension $D$.
For instance, it was 
surprisingly found that the chaotic BKL oscillatory behaviour 
disappears from the generic solution of the vacuum Einstein 
equations in spacetime dimension $D \geq 11$ and is replaced by a 
monotonic Kasner-like power-law behaviour \cite{DHS}. Second, as we said above,
 the generic solution of the 
Einstein-scalar equations also exhibits a non-oscillatory, 
power-law behaviour \cite{BK}, \cite{AR00} (in any dimension \cite{DHRW}).

In superstring theory \cite{GSW,polchinski} there are many massless 
(bosonic) degrees of freedom which can be generically excited near a 
cosmological singularity. They correspond to a high-dimension ($D = 10$ or 
$11$) Kaluza-Klein-type model containing, in addition to Einstein's 
$D$-dimensional gravity, several other fields (scalars, vectors 
and/or forms). In view of the results quoted above, it is a priori 
unclear whether the full field content of superstring theory will 
imply, as generic cosmological solution, a chaotic BKL-like 
behaviour, or a monotonic Kasner-like one. It was found in \cite{DH1,DH2,DH3} 
 that the massless bosonic content of all 
superstring models ($D = 10$ IIA, IIB, I, ${\rm het}_{\rm E}$, ${\rm 
het}_{\rm SO}$), as well as of 
$M$-theory ($D=11$ supergravity), generically implies a chaotic 
BKL-like oscillatory behaviour near
a cosmological singularity. 
[The  analysis  of \cite{DH1,DH2,DH3} applies at
scales large enough to excite all Kaluza-Klein-type modes, but small
enough to be able to neglect the stringy and non-perturbative
massive states.]
It is
the presence of various form fields (e.g. the three form in ${\rm 
SUGRA}_{11}$) which provides the crucial source of this generic 
oscillatory behaviour.

Let us consider a model of the general form
\begin{equation}
S = \int \, d^D \, x \, \sqrt{g} \big[ R(g) - \partial_{\mu} \, 
\varphi \, \partial^{\mu} \, \varphi \,
- \sum_p \ \frac{1}{(p+1)!} \ 
e^{\lambda_p \, \varphi} \, (d \, A_p)^2 \big] \, . \label{eq1}
\end{equation}
Here, the spacetime dimension $D$ is left unspecified. We work (as a 
convenient common formulation) in the Einstein conformal frame, and 
we normalize the kinetic term of the ``dilaton'' $\varphi$ with a 
weight 1 with respect to the Ricci scalar. [Note that this differs of the convention of
Eq.(\ref{1.1}) where there was a factor $1/2$.]
The integer $p \geq 0$ 
labels the various $p$-forms $A_p \equiv A_{\mu_1 \ldots \mu_p}$ 
present in the theory, with field strengths $F_{p+1} \equiv d \, 
A_p$, i.e. $F_{\mu_0 \, \mu_1 \ldots \mu_p} = \partial_{\mu_0} \, 
A_{\mu_1 \ldots \mu_p} \pm p$ permutations. The real parameter 
$\lambda_p$ plays the crucial role of measuring the strength of the 
coupling of the dilaton to the $p$-form $A_p$ (in the Einstein 
frame). When $p=0$,
we assume that $\lambda_0 \not= 0$
(this is the case in type IIB where there is
a second scalar).  
The Einstein metric $g_{\mu \nu}$ is used to lower or raise 
all indices in Eq.~(\ref{eq1}) ($g \equiv -\det \, g_{\mu \nu}$). 
The model (\ref{eq1}) is, as it reads, not quite general enough to 
represent in detail all the superstring actions. Indeed, it lacks 
additional terms involving possible couplings between the form 
fields (e.g. Yang-Mills couplings for $p=1$ multiplets, Chern-Simons 
terms, $(d \, C_2 - C_0 \, d \, B_2)^2$-type terms 
in type IIB). However, it has been verified in all relevant 
cases that these additional terms do not qualitatively modify the 
BKL behaviour to be discussed below. On the other hand, in the case 
of $M$-theory (i.e. $D=11$ SUGRA) the dilaton 
$\varphi$ is absent, and one must cancell 
its contributions to the dynamics.

The leading Kasner-like approximation to the solution of the field 
equations for $g_{\mu \nu}$ and $\varphi$
derived from (\ref{eq1}) is, as usual \cite{BKL}, in the Einstein-frame (see above
for its string-frame counterpart)
\begin{equation}
\label{eq2}
g_{\mu \nu} \, dx^{\mu} \, dx^{\nu} \simeq -dt^2 + \sum_{i=1}^d \, 
t^{2 p_i (x)} \, (\omega^i)^2 \, , \; \;
\varphi \simeq p_{\varphi} \, (x) \, \ln \, t \, +
\psi \, (x) \, ,  
\end{equation}
where $d \equiv D-1$ denotes the spatial dimension and where 
$\omega^i \, (x) = e_j^i \, (x) \, dx^j$ is a time-independent 
$d$-bein. The spatially dependent Kasner exponents $p_i \, (x)$, 
$p_{\varphi} \, (x)$ must satisfy the famous Kasner constraints 
(modified by the presence of the dilaton):
\begin{equation}
p_{\varphi}^2 + \sum_{i=1}^d \ p_i^2 - \left( 
\sum_{i=1}^d \ p_i \right)^2 = 0 \, , \; \;
\sum_{i=1}^d \ p_i = 1 \, . \label{eq3} 
\end{equation}
The set of parameters satisfying Eqs.~(\ref{eq3}) is topologically
a $(d-1)$-dimensional sphere: the ``Kasner sphere''. When the 
dilaton is absent, one must set $p_{\varphi}$ to zero in 
Eqs.(\ref{eq3}). In that case the dimension of the Kasner sphere is 
$d-2 = D-3$.

The approximate solution Eqs.~(\ref{eq2}) is obtained by neglecting 
in the field equations for $g_{\mu \nu}$ and $\varphi$: (i) the 
effect of the spatial derivatives of $g_{\mu \nu}$ and $\varphi$, 
and (ii) the contributions of the various 
$p$-form fields $A_p$. The condition for the ``stability'' of the 
solution (\ref{eq2}), i.e. for the absence of BKL oscillations at $t 
\rightarrow 0$, would be that the inclusion in the field equations of the 
discarded contributions (i) and (ii) (computed within the assumption 
(\ref{eq2})) be fractionally negligible as $t \rightarrow 0$. As 
usual, the fractional effect of the spatial derivatives of $\varphi$ 
is found to be negligible, while the fractional effect (with respect 
to the leading terms, which are $\propto t^{-2}$) of the spatial 
derivatives of the metric, i.e. the quantities $t^2 \, 
\overline{R}_j^i$ (where $\overline{R}_j^i$ denotes the 
$d$-dimensional Ricci tensor) contains, as only ``dangerous terms'' 
when $t \rightarrow 0$ a sum of terms $\propto t^{2 g_{ijk}}$, where 
the {\it gravitational exponents} $g_{ijk}$ ($i \ne j$, $i \ne k$, 
$j \ne k$) are the following combinations of the Kasner exponents 
\cite{DHS}
\begin{equation}
g_{ijk} \, (p) = 2 \, p_i + \sum_{\ell \ne i,j,k} \ p_{\ell} = 1 + 
p_i - p_j - p_k \, . \label{eq4}
\end{equation}
The ``gravitational'' stability condition is that all the exponents 
$g_{ijk} \, (p)$ be positive. In the presence of form fields $A_p$ there 
are additional stability conditions related to the contributions
of the form fields to the Einstein-dilaton equations.
They are obtained by solving, 
\`a la BKL, the $p$-form field equations in the background 
(\ref{eq2}) and then estimating the corresponding ``dangerous'' 
terms in the Einstein field equations. When
neglecting the spatial derivatives in the Maxwell equations in 
first-order form $d \, F = 0$ and $\delta \,
(e^{\lambda_p \, \varphi} \, F) = 0$, where $\delta \equiv * \, d \,*$ 
is the (Hodge) dual of the Cartan differential $d$
and $F_{p+1} = d \, A_p$, one 
finds that both the ``electric'' components $\sqrt g \, e^{\lambda_p 
\, \varphi} \, F^{0 i_1 \ldots i_p}$, and the ``magnetic'' 
components $F_{j_1 \ldots j_{p+1}}$, are constant in time. 
Combining this information with the approximate results 
(\ref{eq2}) one can estimate the fractional effect of the $p$-form 
contributions in the right-hand-side of the $g_{\mu \nu}$- and 
$\varphi$-field equations, i.e. the quantities $t^2 \, T_{(A)0}^0$ 
and $t^2 \, T_{(A)j}^i$ where $T_{(A)\nu}^{\mu}$ denotes the 
stress-energy tensor of the $p$-form. [As usual \cite{BKL} the mixed 
terms $T_{(A)i}^0$, which enter the momentum constraints play a 
rather different role and do not need to be explicitly 
considered.] Finally, one gets as ``dangerous'' terms when $t 
\rightarrow 0$ a sum of ``electric'' contributions $\propto \, t^{2 
\, e_{i_1 \ldots i_p}^{(p)}}$ and of ``magnetic'' ones $\propto \, 
t^{2 \, b_{j_1 \ldots j_{d-p-1}}^{(p)}}$. Here, the {\it electric 
exponents} $e_{i_1 \ldots i_p}^{(p)}$ (where all the indices $i_n$ 
are different) are defined as
\begin{equation}
e_{i_1 \ldots i_p}^{(p)} \, (p) = p_{i_1} + p_{i_2} + \cdots + 
p_{i_p} - \frac{1}{2} \ \lambda_p \, p_{\varphi} \, , \label{eq5}
\end{equation}
while the {\it magnetic exponents} $b_{j_1 \ldots j_{d-p-1}}^{(p)}$ 
(where all the indices $j_n$ are different) are
\begin{equation}
b_{j_1 \ldots j_{d-p-1}}^{(p)} \, (p) = p_{j_1} + p_{j_2} + \cdots + 
p_{j_{d-p-1}} + \frac{1}{2} \, \lambda_p \, p_{\varphi} \, . 
\label{eq6}
\end{equation}
To each $p$-form is associated a (duality-invariant)
double family of ``stability'' 
exponents $e^{(p)}$, $b^{(p)}$. The ``electric'' (respectively
``magnetic'') stability condition is that all the exponents $e^{(p)}$
(respectively, $b^{(p)}$) be positive. 

In \cite{DH1}, it was found that there exists no open region of the Kasner sphere
where all the stability exponents can be simultaneously positive. This showed
that the generic cosmological solution in string theory was of the never-ending
oscillatory BKL type. A deeper understanding of the structure of this generic
solution was then obtained by mapping the dynamics of the scale factors, and of the
dilaton, onto a {\it billiard motion}. 
Let us recall that the
central idea of the BKL approach is that the various points in space approximately decouple as
one approaches a spacelike singularity $(t \rightarrow 0)$. More precisely, the partial
differential equations that control the time evolution of the fields 
can be replaced by ordinary differential equations with
respect to time, with coefficients that are (relatively) slowly varying in space and time.
The details of how this is done are explained in \cite{BKL,DH1,DHN}.
Let us review the main result of \cite{DH3}, namely the fact 
 that the evolution of the scale factors and the dilaton at each
spatial point can be be viewed as a billiard motion in some simplices in hyperbolic space $H^9$, which
have remarkable connections with hyperbolic Kac-Moody algebras of rank $10$. 

To see this we generalize the previous Kasner-like solution by expressing it in terms
of some local scale factors, $a_i$, without assuming that these scale factors
behave as powers of the proper time (as was done in (\ref{eq2}) which had assumed
$a_i \propto t^{p_i}$). In other words, we now write the metric (in
either the Einstein frame or the string frame) as $g_{\mu \nu} \, dx^{\mu} \, dx^{\nu} = -N^2
(dx^0)^2 + \sum_{i=1}^{d} \ a_i^2 \, (\omega^i)^2$, where $d \equiv D-1$ denotes the spatial
dimension, and where, as above,
 $\omega^i (x) = e_j^i (x) \, dx^j$ is a $d$-bein whose time-dependence is
neglected compared to that of the local scale factors $a_i$. 
Instead of working with the 9 variables $a_i$ , and the dilaton $\varphi$,  
it is convenient to introduce the following set of
10 field variables: $\beta^{\mu}, \;  \mu = 1,\ldots, 10$, with, in the superstring 
(Einstein-frame) case,
$\beta^i \equiv -\ln \, a_i$ ($i = 1, \ldots, 9$), and $\beta^{10} \equiv -\varphi$ where $\varphi$ is 
the Einstein-frame dilaton. 
[In $M$-theory there is no dilaton but $\mu \equiv i =  1, \ldots, 10$. In the string
frame, we define $\beta_S^0 \equiv -\ln ( \sqrt {g_S} e^{-2 \Phi})$
and label $\mu = 0, \ldots, 9$.] 

We consider the evolution near a past (big-bang) or future (big-crunch) 
spacelike singularity located at $t=0$, where $t$ is the
proper time from the singularity. In the gauge $N = -\sqrt g$ (where $g$ is the
determinant of the Einstein-frame spatial metric), 
i.e. in terms of the new time variable $ d\tau = - dt / \sqrt
g$, the action (per unit comoving volume) describing the asymptotic dynamics of $\beta^{\mu}$ as
$t \rightarrow 0^+$ or $\tau \rightarrow + \infty$ has the form
\begin{equation}
S = \int d \tau \left[ G_{\mu \nu} \, \frac{d \beta^{\mu}}{d\tau} \ \frac{%
d\beta^{\nu}}{d\tau} - V (\beta^{\mu}) \right] \, ,  \label{eq1'}
\end{equation}
\begin{equation}
V (\beta) \simeq \sum_A \, C_A \, e^{ - 2 w_A (\beta)} \, .  \label{eq2'}
\end{equation}
In addition, the time reparametrization invariance (i.e. the equation of motion of $N$ in a
general gauge) imposes the usual ``zero-energy" constraint
$E = G_{\mu \nu} (d\beta^{\mu}/{d\tau}) (d\beta^{\nu}/{d\tau}) +
V (\beta^{\mu}) = 0$.
The metric $G_{\mu \nu}$ in field-space is a 10-dimensional metric of Lorentzian signature $-++
\cdots +$. Its explicit expression depends on the model and the choice of variables. In
$M$-theory, $ G_{\mu \nu}^M \, d\beta_M^{\mu} \, d\beta_M^{\nu} = \sum_{\mu=1}^{10} \,
(d\beta_M^{\mu})^2 - \left( \sum_{\mu=1}^{10} \, d\beta_M^{\mu} \right)^2$,
while in the string models, 
 $ G_{\mu \nu}^S \, d\beta_S^{\mu} \, d\beta_S^{\nu} = \sum_{i=1}^{9} \, (d\beta_S^i)^2
- (d\beta_S^0)^2$ in the string frame.
 Each exponential term, labelled by $A$, in the potential
$V (\beta^{\mu}),$ Eq.~(\ref {eq2'}),
represents the effect, on the evolution of $(g_{\mu \nu} , \varphi)$, of either (i) the spatial
curvature of $g_{ij}$ (``gravitational walls''), (ii) the energy density of some electric-type
components of some $p$-form $A_{\mu_1 \ldots \mu_p}$ (``electric $p$-form
wall''), or (iii) the energy density of some magnetic-type components of $%
A_{\mu_1 \ldots \mu_p}$ (``magnetic $p$-form wall''). The coefficients $C_A$ are all found to be
positive, so that all the exponential walls in Eq.~(\ref {eq2'}) are repulsive. The $C_A$'s vary in
space and time, but we neglect their variation compared to the asymptotic effect of $w_A (\beta)$
discussed below. Each exponent $ - 2 \, w_A (\beta)$ appearing in Eq.~(\ref{eq2'}) is a {\it linear
form} in the $\beta^{\mu} : w_A (\beta) = w_{A\mu} \, \beta^{\mu} $.  
The wall forms $w_A (\beta)$ are exactly the same linear forms as the ``stability exponents''
which appeared above; one just need to replace the variables $p_i$ by $\beta^{\mu}$. For
instance, one of the ``electric'' wall forms for a $3$-form coupled with $\lambda = 0$  is
$w_{123}(\beta) = e^{(3)}_{123}(\beta)=  \beta^1 + \beta^2 + \beta^3$.
The complete list of ``wall
forms'' $w_A (\beta)$, was given in \cite{DH1} for each string model.  The number of walls is
enormous, typically of the order of $700$.

At this stage, one sees that the $\tau$-time dynamics of the variables $ \beta^{\mu}$ is described
by a Toda-like system in a Lorentzian space, with a zero-energy constraint. But it seems
daunting to have to deal with $\sim 700$ exponential walls! However, the problem can be greatly
simplified because many of the walls turn out to be asymptotically irrelevant. To see this, it
is useful to project the motion of the variables $\beta^{\mu}$ onto
the 9-dimensional hyperbolic space $H^9$ (with curvature $-1$). This can be done because the
motion of $\beta^{\mu}$ is always time-like, so that, starting (in our units) from the origin, it
will remain within the 10-dimensional Lorentzian light cone of $G_{\mu \nu}$. This follows from
the energy constraint and the positivity of $V$ .  With our definitions,
 the evolution occurs in the {\it future} light-cone. The
projection to $H^9$ is performed by decomposing the motion of $\beta^{\mu}$ into its radial and
angular parts (see \cite{Chitre,Misner} and the generalization \cite{Melnikov}). One writes
 $\beta^{\mu} =  + \rho \, \gamma^{\mu}$ with $\rho^2 \equiv - G_{\mu \nu} \,
\beta^{\mu} \, \beta^{\nu}$, $\rho >0$ and $G_{\mu \nu} \, \gamma^{\mu} \, \gamma^{\nu} = -1$ (so
that $\gamma^{\mu}$ runs over $H^9$, realized as the {\it future}, unit hyperboloid) and one introduces
a new evolution parameter: $dT =  k \, d\tau / \rho^2$. The action (\ref{eq1'}) becomes
\begin{equation}
S = k \int dT \left[ - \left( \frac{d \, \ln \rho}{dT} \right)^2 + \left(
\frac{d \mbox{\boldmath$\gamma$}}{dT} \right)^2 - V_T (\rho , %
\mbox{\boldmath$\gamma$}) \right]  \label{eq7}
\end{equation}
where $d \mbox{\boldmath$\gamma$}^2 = G_{\mu \nu} \, d \gamma^{\mu} \, d \gamma^{\nu}$ is the
metric on $H^9$, and where $V_T = k^{-2} \, \rho^2 \, V = \sum_A \ k^{-2} \, C_A
\, \rho^2 \, \exp (-2 \, \rho \, w_A (\mbox{\boldmath$\gamma$}))$. When $t \rightarrow
0^+$, i.e. $\rho \rightarrow +\infty$, the transformed potential $V_T (\rho ,
\mbox{\boldmath$\gamma$})$ becomes sharper and sharper and reduces in the limit to a set of
$\rho$-independent impenetrable walls located at $w_A(\mbox{\boldmath$\gamma$}) = 0$
 on the unit hyperboloid (i.e. $V_T = 0$ when $w_A(\mbox{\boldmath$\gamma$}) > 0$,
and $V_T = +\infty$ when $w_A(\mbox{\boldmath$\gamma$}) < 0$). 
In this limit, $d \, \ln \rho / dT$ becomes constant, and one can
choose the constant $k$ so that $d \, \ln \rho / dT = 1 $.  The (approximately) linear motion 
of $\beta^{\mu}(\tau)$ between two ``collisions'' with the original multi-exponential potential
 $V (\beta^{\mu})$ is thereby mapped onto a geodesic motion of $\mbox{\boldmath$\gamma$} (T)$
  on $H^9$, interrupted by specular collisions on sharp hyperplanar walls.
   This motion has unit velocity $(d \mbox{\boldmath$\gamma$}
/ dT)^2 = 1$ because of the energy constraint.  In
terms of the original variables $\beta^{\mu}$, the motion is confined to the convex domain 
(a cone in
a 10-dimensional Minkowski space) defined by the intersection of the {\it positive} sides of all the
wall hyperplanes $w_A (\beta) = 0$ and of the interior of the
 future light-cone $G_{\mu \nu} \, \beta^{\mu} \,\beta^{\nu} = 0$.

A further, useful simplification is obtained by quotienting the dynamics of $\beta^{\mu}$ by the
natural permutation symmetries inherent in the problem, which correspond to ``large
diffeomorphisms" exchanging the various proper directions of expansion and the corresponding
scale factors. The natural configuration
space is therefore ${\fam\bbfam R}^d / {\rm S}_d$, which can be parametrized by the {\it ordered
multiplets} $ \beta^1 \leq \beta^2 \leq \cdots \leq \beta^d$. 
This kinematical 
quotienting is standard in most investigations of the BKL oscillations \cite{BKL}
and can be implemented in ${\fam\bbfam R}^d$ by introducing further sharp walls located at
$\beta^i = \beta^{i+1}$. These ``permutation walls'' have been recently derived from
a direct dynamical analysis based on the Iwasawa decomposition of the metric \cite{DHN}.
 Finally the
dynamics of the models is equivalent, at each spatial point, to a hyperbolic billiard problem. The
specific shape of this model-dependent billiard is determined by the original walls and 
the permutation walls. Only the ``innermost'' walls (those which are not ``hidden'' behind others)
are relevant.

The final results of the analysis of the innermost walls
 are remarkably simple. Instead of the ${\cal O} (700)$
original walls it was found, in all cases, that there are only 10 relevant walls. In fact, the seven
string theories M, IIA, IIB, I, HO, HE and the closed bosonic string in $D=10$ \cite{boso}, split
into {\it three} separate blocks of theories, corresponding to three distinct billiards.
 The first block (with 2 SUSY's in $D = 10$) is ${\cal B}_2 = \{$M, IIA, IIB$\}$ and
its ten walls are (in the natural variables of $M$-theory $\beta^{\mu} = \beta_M^{\mu}$),
\begin{eqnarray}
{\cal B}_2: w_i^{[2]} (\beta) &=& - \beta^i + \beta^{i+1} ( i = 1, \ldots , 9), \nonumber \\
w_{10}^{[2]} (\beta) &=& \beta^1 + \beta^2 + \beta^3.
\end{eqnarray}
The second block is ${\cal B}_1 = \{$I, HO, HE$\}$ and its ten walls read (when written in terms
of the string-frame variables of the heterotic theory $\alpha^i = \beta_S^i$, $\alpha^0 =
\beta_S^0$; see Eqs.(\ref{strein}))
\begin{eqnarray}
{\cal B}_1: w_1^{[1]} (\alpha) &=& \alpha^1, \; w_i^{[1]} (\alpha) = - \alpha^{i-1} + \alpha^i (i
= 2
, \ldots , 9), \nonumber \\
w_{10}^{[1]} (\alpha) &=& \alpha^0 - \alpha^7 - \alpha^8 - \alpha^9.
\end{eqnarray}
The third block is simply ${\cal B}_0 = \{ D = 10$ closed bosonic$\}$ and its ten walls read (in string
variables) 
\begin{eqnarray}
{\cal B}_0: w_1^{[0]} (\alpha) &=& \alpha^1 + \alpha^2, \;  w_i^{[0]} (\alpha) = -
\alpha^{i-1} + \alpha^i (i = 2, \ldots , 9), \nonumber \\
  w_{10}^{[0]} (\alpha) &=& \alpha^0 - \alpha^7 -\alpha^8 - \alpha^9.
\end{eqnarray}
 In all cases, these walls define a simplex of $H^9$ which is non-compact but
of finite volume, and which has remarkable symmetry properties. 

\begin{figure}[ht]
\hspace{30mm}
\psfig{figure=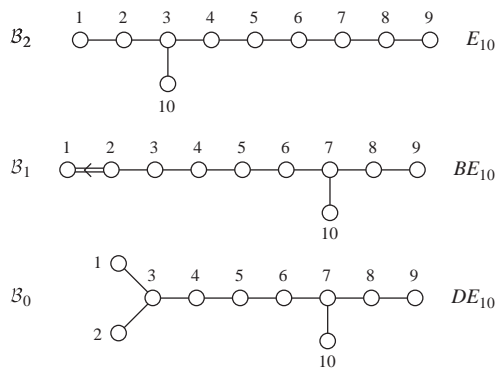}
\caption{Dynkin diagrams defined (for each $n = 2,1,0$) by the ten wall forms
 $w_i^{[n]}(\beta^{\mu}), i = 1,\ldots,10$ that determine the billiard dynamics, near a
cosmological singularity, of the
three blocks of theories ${\cal B}_2 = \{$M, IIA, IIB$\}$, ${\cal B}_1 = \{$I, HO, HE$\}$ and
${\cal B}_0 = \{D = 10$ closed bosonic$\}$. The node labels $ 1,\ldots,10$ correspond to
the form label $i$ used in the text.}
\label{fig1}
\end{figure}

The most economical way to describe the geometry of the simplices is through their Coxeter
diagrams. This diagram encodes the angles between the faces and is obtained by computing the Gram
matrix of the scalar products between the unit normals to the faces, say $\Gamma_{ij}^{[n]} \equiv
\widehat{w}_i^{[n]} \cdot \widehat{w}_j^{[n]}$ where $\widehat{w}_i \equiv w_i / \sqrt{w_i \cdot
w_i}$, $i = 1 , \ldots , 10 $ labels the forms defining the (hyperplanar) faces of a simplex, and
the dot denotes the scalar product (between co-vectors) induced by the metric $ G_{\mu \nu} : w_i
\cdot w_j \equiv G^{\mu \nu} \, w_{i\mu} \, w_{j\nu}$ for $ w_i (\beta) = w_{i\mu} \,
\beta^{\mu}$. This Gram matrix does not depend on the normalization of the forms $w_i$ but
actually, all the wall forms $w_i$ listed above are normalized in
a natural way, i.e. have a natural length.
This is clear for the forms which are directly associated with dynamical walls in $D = 10$ or 11,
but this can also be extended to all the permutation-symmetry walls because
 they appear as dynamical walls after dimensional reduction \cite{DH3,DHN}. When the wall
forms are normalized accordingly (i.e. such that $V_i^{{\rm dynamical}} \propto \exp (-2 \, w_i
(\beta)$), they all have a squared length $w_i^{[n]} \cdot w_i^{[n]} = 2$, {\it except} for $
w_1^{[1]} \cdot w_1^{[1]} = 1$ in the ${\cal B}_1$ block. We can then compute
 the ``Cartan matrix'', $ a_{ij}^{[n]} \equiv 2 \, w_i^{[n]} \cdot
w_j^{[n]} / w_i^{[n]} \cdot w_i^{[n]}$, and the corresponding Dynkin diagram. 
One finds the diagrams given in Fig.~\ref{fig1}.

  The corresponding Coxeter diagrams are obtained
from the Dynkin diagrams by forgetting about the norms of the wall forms, i.e., by deleting the
arrow in $BE_{10}$. As can be seen from the figure, the Dynkin diagrams associated with the
billiards turn out to be the Dynkin diagrams of the following rank-$10$ hyperbolic Kac-Moody
algebras (see \cite {Kac}): E$_{10}$, BE$_{10}$ and DE$_{10}$ (for ${\cal B}_2$, ${\cal B}_1$ and
${\cal B}_0$, respectively). It is remarkable that the three billiards exhaust the only three
possible simplex Coxeter diagrams on $H^9$ with {\it discrete} associated Coxeter group (and this
is the highest dimension where such simplices exist) \cite {Vinberg}.
This analysis suggests to identify the 10 {\it wall forms} $w_i^{[n]} (\beta)$, $i = 1, \ldots , 10$ of
the billiards ${\cal B}_2$, ${\cal B}_1$ and ${\cal B}_0$ with a basis of {\it simple roots} of the
hyperbolic Kac-Moody algebras E$_{10}$, BE$_{10}$ and DE$_{10}$, so that the cosmological {\it billiard}
can be identified with a {\it fundamental Weyl chamber} of these algebras. Note also that
 the 10 dynamical variables
$ {\it \beta^{\mu}}$, $\mu = 1 , \ldots , 10$, can be considered as parametrizing a generic vector in the
{\it Cartan subalgebra} of these algebras. 

It was conjectured some time ago \cite{Julia} that E$_{10}$
should be, in some sense, the symmetry group of SUGRA$_{11}$ reduced to one dimension 
(and that DE$_{10}$ be that of type I SUGRA$_{10}$, which has the same bosonic spectrum as
the bosonic string). Our
results, which indeed concern  the one-dimensional ``reduction''\footnote{Note again that the analysis above
concerns generic inhomogeneous solutions depending upon $D$ variables. The strict one-dimensional
reduction (one variable only) of M-theory has also been considered, and has been shown to still
contain the Weyl group of E$_{10}$ \cite{DH2}.}, \`a la BKL, of $M$/string theories
 exhibit a clear sense in which E$_{10}$ lies behind the one-dimensional
evolution of the block ${\cal B}_2$ of theories: their asymptotic cosmological evolution as $ t
\rightarrow 0$ is a billiard motion, and the group of reflections in the walls of this billiard is
nothing else than the {\it Weyl group} of E$_{10}$ (i.e. the group of reflections in the
hyperplanes corresponding to the roots of E$_{10}$, which can be generated by the 10 simple roots
of its Dynkin diagram). 
It is intriguing -- and, to our knowledge, unanticipated 
(see, however, \cite{CJLP})-- that the
cosmological evolution of the second block of theories, ${\cal B}_1 = \{$I, HO, HE$\}$, be
described by {\it another} remarkable billiard, whose group of reflections is the Weyl group of
BE$_{10}$. The root lattices of E$_{10}$ and BE$_{10}$ exhaust the only two possible
unimodular even and odd Lorentzian 10-dimensional lattices \cite {Kac}.


A first consequence of the exceptional properties of the billiards concerns the nature of the
cosmological oscillatory behaviour. They lead to a direct technical proof that these oscillations,
for all three blocks, are chaotic in a mathematically well-defined sense. This is done
by reformulating, in a standard manner, the billiard dynamics as an equivalent collision-free
geodesic motion on a hyperbolic, finite-volume {\it manifold} (without boundary)
 ${\cal M}$ obtained by quotienting $H^9$
by an appropriate torsion-free discrete group. 
These geodesic motions define the ``most chaotic'' type of dynamical systems.
They are Anosov flows \cite{Anosov}, which imply, in particular, that they are ``mixing''. In
principle, one could (at least numerically) compute their largest, positive Lyapunov exponent, say
$\lambda^{[n]}$, and their (positive) Kolmogorov-Sinai entropy, say $h^{[n]}$. As we work on a
manifold with curvature normalized to $-1$, and walls given in terms of equations containing only
numbers of order unity, these quantities will also be of order unity. Furthermore, the two Coxeter
groups of E$_{10}$ and BE$_{10}$ are the only reflective {\it arithmetic} groups in 
$H^9$ \cite{Vinberg} so that the chaotic
motion in the fundamental simplices of E$_{10}$ and BE$_{10}$ will be of the exceptional
``arithmetical'' type \cite{BGGS}. We therefore expect that
 the quantum motion on these two billiards, and in particular the spectrum of
the Laplacian operator, exhibits exceptional features (Poisson statistics of
level-spacing,$\ldots$), linked to the existence of a Hecke
algebra of mutually commuting, conserved operators. Another (related) remarkable feature of the
billiard motions for all these blocks is their link, pointed out above,
 with Toda systems. This fact is probably quite significant, both classically and 
 quantum mechanically, because
Toda systems whose walls are given in terms of the simple roots of a Lie algebra enjoy 
remarkable properties. We leave to future work a study of our Toda systems which involve
infinite-dimensional hyperbolic Lie algebras.

The discovery that the chaotic behaviour of the generic cosmological solution of superstring
effective Lagrangians was rooted in the fundamental Weyl chamber of some underlying hyperbolic
Kac-Moody algebra prompted us to reexamine the case of pure gravity \cite{DHJN}. It was found
that the same remarquable connection applies to pure gravity in any dimension 
$D \equiv d + 1 \geq 4$. The relevant Kac-Moody algebra in this case is $AE_d$. It was
also found that the disappearance of chaos in pure gravity models when $ D \geq 11$ dimensions
\cite{DHS} becomes linked to the fact that the Kac-Moody algebra $AE_d$ is no longer ``hyperbolic''
for $d \geq 10$ \cite{DHJN}.

The present investigation a priori concerned only the ``low-energy'' $(E \ll
(\alpha^{\prime})^{-1/2})$, classical cosmological behaviour of string
theories. In fact, if (when going toward the singularity) one starts at some
``initial'' time $t_0 \sim (d\beta / dt)_0^{-1}$ and
 insists on limiting the application of our results
to time scales $\vert t \vert \lower.5ex\hbox{$\; \buildrel > \over \sim \;$}
(\alpha^{\prime})^{1/2} \equiv t_s$, the total number of ``oscillations'',
i.e. the number of collisions on the walls of our billiard will be finite,
and will not be very large. The results above show that the number of
collisions between $t_0$ and $t \rightarrow 0$ is of order $N_{{\rm %
coll}} \sim \ln \tau \sim \ln (\ln (t_0/t))$. This is only $N_{{\rm coll}}
\sim 5$ if $t_0$ corresponds to the present Hubble scale and $t$ to the
string scale $t_s$. However, the
strongly mixing properties of geodesic motion on hyperbolic spaces make it
large enough for churning up the fabric of spacetime and transforming any,
non particularly homogeneous at time $t_0$, patch of space into a turbulent
foam at $t = t_{s}$. Indeed, the mere fact that the walls
associated with the spatial curvature and the form fields repeatedly rise up
(during the collisions) to the same level as the ``time'' curvature terms $%
\sim t^{-2}$, means that the spatial inhomogeneities at $t \sim t_s$ will also be of order
$t_s^{-2}$, corresponding to a string scale foam.

Our results on the ${\cal B}_2$ theories probably involve a deep (and not a priori evident)
connection with those of Ref.~\cite{BFM} on the structure of the moduli space of $M$-theory
compactified on the ten torus $T^{10}$, with vanishing 3-form potential (see also
\cite{obers}). In both cases the Weyl
group of E$_{10}$ appears. In our case it is (partly) {\it dynamically} realized as reflections in
the walls of a billiard, while in Ref.~\cite{BFM} it is {\it kinematically} realized as a symmetry
group of the moduli space of compactifications preserving the maximal number of supersymmetries.
In particular, the crucial E-type node of the Dynkin diagram of E$_{10}$ (Fig.~\ref{fig1}) comes,
in our study and in the case of $M$-theory, from the wall form $w_{10}^{[2]} (\beta) = \beta_M^1 +
\beta_M^2 + \beta_M^3$ associated with the electric energy of the 3-form. By contrast, in
\cite{BFM} the 3-form is set to zero, and the reflection in $w_{10}^{[2]}$ comes from the 2/5
duality transformation (which is a double $T$ duality in type II theories), which exchanges (in
$M$-theory) the 2-brane and the 5-brane. As we emphasized above, dimensional reduction transforms
kinematical (permutation) walls into dynamical ones. This suggests that there is no difference of
nature between our walls, and that, viewed from a higher standpoint (12-dimension ?), they would
all look kinematical, as they are in \cite{BFM}. By analogy, our findings for the ${\cal B}_1$
theories suggest that the Weyl group of BE$_{10}$ is a symmetry group of the moduli space of
 $T^{9}$ compactifications of $\{$I, HO, HE$\}$.

Perhaps the most interesting aspect of the above ``billiard'' analysis is to provide hints for a scenario of
vacuum selection in string cosmology. If we heuristically extend our (classical, low-energy,
tree-level) results to the quantum, stringy $(t \sim t_s)$ and/or strongly coupled $(g_s \sim 1)$
regime, we are led to conjecture that the initial state of the universe is equivalent to the
quantum motion in a certain {\it finite volume chaotic} billiard. This billiard is (as in a hall
of mirrors game) the fundamental polytope of a discrete symmetry group which contains, as
subgroups, the Weyl groups of both E$_{10}$ and BE$_{10}$ \cite{footnote}. We are here assuming
that there is (for finite spatial volume universes) a non-zero transition amplitude between the
moduli spaces of the two blocks of superstring ``theories'' (viewed as ``states'' of an underlying
theory). If we had a description of the resulting combined moduli space (orbifolded by its
discrete symmetry group) we might even consider as most probable initial state of the
universe the fundamental mode of the combined billiard, though this does not seem crucial
for vacuum
selection purposes. This picture is a generalization of the picture of Ref.~%
\cite{HM} (as well as, in some sense, of the ``stochastic'' PBB picture reviewed above)
and, like the latter, might solve the problem of cosmological
vacuum selection in allowing the initial state to have a finite probability
of exploring the subregions of moduli space which have a chance of inflating
and evolving into our present universe.

\vspace{5mm}

\noindent {\bf Acknowledgments}:
I would like to congratulate Marc Henneaux (Francqui Prize, 2000)
for a well deserved recognition, and to
thank him for a most pleasant, enriching and 
fruitful collaboration. I wish also to thank Marc Henneaux and Alexander
Sevrin for organizing an extremely stimulating meeting, and the Francqui Foundation 
for sponsoring in such an elegant and generous manner this very timely colloquium.


\begin{thebibliography}{99}
\bibitem{boso} We included for the sake of comparison the bosonic string model in $D=10$.
Though $D = 10$ is not the critical dimension of the quantum bosonic string, it is
the critical dimension above which the never ending oscillations in its cosmological evolution
disappear \cite{DH2}.
\end{thebibliography}
\end{document}